\documentclass[prl,twocolumn,showpacs,preprintnumbers,amsmath,amssymb,superscriptaddress]{revtex4}

\usepackage{graphicx}% Include figure files
\usepackage{dcolumn}% Align table columns on decimal point
\usepackage{bm}% bold math
\usepackage{color}
\usepackage{epsfig,float,afterpage,amssymb,wrapfig,psfrag}

\DeclareMathAlphabet{\bi}{OML}{cmm}{b}{it}

\newcommand{\be}{\begin{equation}}
\newcommand{\bea}{\begin{eqnarray}}

\newcommand{\ee}{\end{equation}}
\newcommand{\eea}{\end{eqnarray}}

\newcommand{\kk}{\bi{k}}

\newcommand{\ra}{\rightarrow}

\newcommand{\w}{\omega}

\begin{document}
\title{Critical Kondo destruction in a pseudogap Anderson model:\\ scaling and relaxational dynamics}
%of quantum critical point}
\author{Matthew T. Glossop}
\affiliation{Department of Physics \& Astronomy, Rice University,
Houston, Texas, 77005, USA}
\author{Stefan Kirchner}
\affiliation{Max Planck Institute for the Physics of Complex Systems,
01187 Dresden, Germany}
\affiliation{Max Planck Institute for Chemical Physics of Solids,
01187 Dresden, Germany}
\author{J.~H.~Pixley}
\affiliation{Department of Physics \& Astronomy, Rice University,
Houston, Texas, 77005, USA}
\author{Qimiao Si}
\affiliation{Department of Physics \& Astronomy, Rice University,
Houston, Texas, 77005, USA}

\date{\today}

\begin{abstract}
We study the pseudogap Anderson model as a prototype system for
critical Kondo destruction. We obtain finite-temperature ($T$) scaling 
functions near its quantum-critical point, by using 
a continuous-time quantum Monte Carlo method and also considering 
a dynamical large-N limit.
We are able to determine the behavior of the scaling functions 
in the typically difficult to access quantum-relaxational
regime ($\hbar \omega < k_{B} T$), and conclude that the relaxation rates for both
the spin and single-particle excitations are linear in temperature.
We discuss the implications of these results for the quantum-critical phenomena in
heavy fermion metals.

\end{abstract}

% PACS:
%   71.10.Hf   Non-Fermi-liquid ground states, electron phase diagrams and
%              phase transitions in model systems
%   71.27.+a   Strongly correlated electron systems; heavy fermions
%   75.20.Hr   Local moment in compounds and alloys; Kondo effect, valence
%              fluctuations, heavy fermions (see also 72.15.Qm Scattering
%              mechanisms and Kondo effect)
%   05.10.Cc   Renormalization group methods

\maketitle

Continuous zero temperature phase transitions in strongly correlated 
electronic and atomic models
have attracted considerable attention as a new
paradigm for addressing
the universal features of correlated quantum systems~\cite{Natphys.08}.  Quantum criticality links two nearby phases and determines 
the physical properties in a large range of temperature and control parameters,
the quantum-critical region, that fans
out from the quantum-critical point (QCP). 
This paradigm is especially pertinent to the understanding of intermetallic rare earth compounds.
The phase diagram of these heavy fermion metals close to the border of antiferromagnetism
features a QCP, but the associated quantum-critical properties are highly unusual
when viewed from the standard description based on Landau's notion of order-parameter 
fluctuations~\cite{SiSteglich.10}.
Especially, inelastic neutron-scattering measurements have shown that
the dynamical spin susceptibility in the quantum-critical regime features
a linear-in-$T$ spin relaxation 
rate and satisfies a frequency over temperature ($\omega/T$) scaling~\cite{Schroeder.00}.
Very recently, Hall-effect measurements
have indicated that
the single-particle
relaxation rate in the quantum-critical regime is also linear in $T$~\cite{Friedemann.09}.

These dynamical scaling and relaxational properties provide important clues to the 
nature of the heavy-fermion QCP. Yet, theoretically, such real-frequency behavior
is difficult to study.
% and no single analytical
%or numerical scheme works
%in generic cases.
Two regimes need to be distinguished: 
the quantum coherent ($\omega> T$)
and quantum-relaxational ($\omega< T$) regimes~\cite{Damle.97} ($\hbar$ and $k_{B}$ are set to 1). 
Calculation
methods (such as 
Monte Carlo simulations) typically work in the imaginary-time domain,
and the nonzero Matsubara frequencies
($\omega_n$) are necessarily in
the $|\omega_n|/T>1$ regime.
Extracting the behavior at real frequencies requires an analytical continuation,
which is in general a numerically ill-conditioned
procedure.
The numerical renormalization group operates on the real-frequency axis,
but it is not reliable
for
the quantum-relaxational regime at nonzero temperatures.

In this Letter, we address %this issue through
the dynamical and relaxational properties of the particle-hole
symmetric pseudogap Anderson model in both frequency regimes. Our motivations to study this model are 
multifold.
In local quantum criticality for heavy-fermion metals,
the   critical destruction of the Kondo
effect~\cite{Si.01,Coleman.01,Loehneysen.07,Gegenwart.08}  is local in space, and the resulting interacting critical modes
are manifested in local correlators which can be studied in quantum-impurity problems.
The pseudogap Anderson model is the simplest impurity problem that contains the 
physics of critical Kondo destruction; it is well known that varying the Kondo coupling yields a QCP
~\cite{Withoff.90,Buxton.98,Ingersent.02,Vojta.01,Glossop.03,Glossop.05,Fritz.06},
which separates a Kondo-screened Fermi-liquid phase from a Kondo-destroyed local-moment phase.
However, a proper understanding of the  dynamical scaling at finite temperatures and
the associated  relaxational behavior is not yet available  
even in this simplest model. 
%See below for comparison of our results with prior
%studies~\cite{Ingersent.02,Glossop.05,Fritz.06}.)
Furthermore, the pseudogap Anderson/Kondo model 
is relevant in a number of realistic physical settings.
It has been  invoked in the context of non magnetic impurities in cuprate 
superconductors~\cite{Vojta.01a}.  It has also been shown that a judicious tuning of a
double quantum-dot system can produce a pseudogap in the effective density 
of states~\cite{DiasdaSilva.06}. 
In disordered metals, a novel phase has been attributed to the occurrence of 
local pseudogaps near the Fermi energy at local-moment sites~\cite{Zhuravlev.07}. 
Finally, the pseudogap Kondo model is the appropriate model to
describe point defects in graphene~\cite{Chen.11}.

We study the model by using a continuous-time quantum
 Monte Carlo approach (CT-QMC)~\cite{Werner.06}.
We determine the full scaling functions at real frequencies and finite 
temperatures for both the dynamical spin susceptibility and single-electron Green's function.
We 
%are able to do so 
achieve this by 
taking advantage of %the
insights gained from exact calculations at real frequencies and finite temperatures in a dynamical large-$N$ limit of the model. 
% and showing that 
The results
in the large-$N$ limit
motivate us to analyze
%apply these results to 
the imaginary-time correlators 
in the physical $N=2$ model in a way that uncovers
%show they have
 the form of a boundary conformally-invariant fixed point.
 The latter, in turn,
can readily be analytically-continued to real
frequency at finite temperatures.
We establish that both
the dynamical spin susceptibility and single-electron Green's function
display an $\omega/T$-scaling
and contain a linear-in-$T$ relaxation rate.
As a by-product, we show that the CT-QMC approach,
which is based on a high-temperature expansion,
%near a QCP;
%whether this method 
can reach low-enough temperatures
with enough accuracy to resolve quantum-critical features.
%has,
%is an interesting question that has,                                              
%to our knowledge, not been addressed before.

%%%%%%%%%%%%%%%%%%%%%%%%%%%%%%%%%%%%%
{\it Pseudogap Kondo model in a dynamical large-N limit:~}
To set the stage for the CT-QMC study, 
we start with the SU(N)$\times$SU(M) Kondo model~\cite{Parcollet.98} in the presence of 
a pseudogap in the limit of large N and M.
In what follows, we set $\hbar=k_B=1$.
The Hamiltonian is
\begin{eqnarray}
{\cal H}_{\text{PKM}} &=&
({J_K}/{N})
\sum_{\alpha}{\bf S}
\cdot {\bf s}_{\alpha}
+ \sum_{p,\alpha,\sigma} E_{p}~c_{p \alpha
\sigma}^{\dagger} c_{p \alpha \sigma} .
\label{HlargeN}
\end{eqnarray}
Here,
the spin and channel indices
are $\sigma = 1, \ldots, N$
and
$\alpha=1, \ldots, M$, respectively.
The conduction electron density of states
takes the form:
\begin{eqnarray}
\rho(\omega)=\sum_p \delta(E_p-\omega)=
\rho_0 |\omega/D|^r\Theta(D-|\omega|) ,
\label{DOS}
\end{eqnarray}
with $2D$ being the bandwidth.
That this limit has a 
nontrivial QCP can be seen through
the particular form of 
the perturbative (in $r$) renormalization group equation~\cite{Withoff.90}.
In the limit 
of large N and M, the renormalization group beta function becomes 
$
%\begin{equation}
%\label{largeNRG}
\beta(j)=-j(r-j+\kappa j^2),
%\end{equation}
$ with $j=J_K/D$ and $\kappa=M/N$~\cite{Zhu.04}.
This establishes that the QCP  survives the large-N limit and
can be accessed perturbatively. 
%Moreover, Eq.~(\ref{largeNRG}), 
To order $r$, the large-N beta function is identical to its
N$=2$ counterpart~\cite{Withoff.90,Ingersent.02} 
suggesting 
that the universal critical scaling properties 
of the $N=2$ QCP
are preserved by taking the large-N limit.
In this limit, the local degrees of freedom are
expressed in terms of pseudofermions $f_{\sigma}$ and a bosonic
decoupling field $B_{\alpha}$, where
$S_{\sigma,\sigma^{\prime}}=f^{\dagger}_{\sigma}f^{}_{\sigma^{\prime}}
-\delta_{\sigma,\sigma^{\prime}}Q/N$,
and $Q$ is related to the chosen irreducible representation
of SU(N)~\cite{Parcollet.98,Cox.93}.
The large-$N$ equations are
\begin{eqnarray}
\Sigma_B(\tau) &=& - {\cal G}_{0}(\tau) G_f(-\tau);~
\Sigma_f(\tau)= \kappa {\cal G}_{0}(\tau) G_B(\tau); \nonumber \\
%\end{eqnarray}
%together with
%\begin{eqnarray}
~~G_B^{-1}( i\nu_n) &=&
1/{J_K} - \Sigma_B( i\nu_n); \nonumber \\
~~G_f^{-1}(i\omega_n)& =& i\omega_n - \lambda -
\Sigma_f(i\omega_n);
\label{NCA}
%\nonumber \\
%G_f(\tau = 0^{-}) &=& ({1}/{\beta}) \sum_{i\omega_n}
%G_f (i\omega_n) {\rm e}^{i\omega_n 0^+} = {1}/{2}
\end{eqnarray}
together with a constraint $G_f(\tau\rightarrow 0^{-})=Q/N$~\cite{Parcollet.98}.
Here, $\lambda$ is a Lagrangian multiplier 
enforcing the constraint and
${\cal G}_0 = - \langle T_{\tau} c_{\sigma\alpha}(\tau)
c_{\sigma\alpha}^{\dagger}(0) \rangle _0$
is the noninteracting Green's function~\cite{Vojta.01}.

\begin{figure}[t!]
\begin{center}
\includegraphics[width=0.5\textwidth]{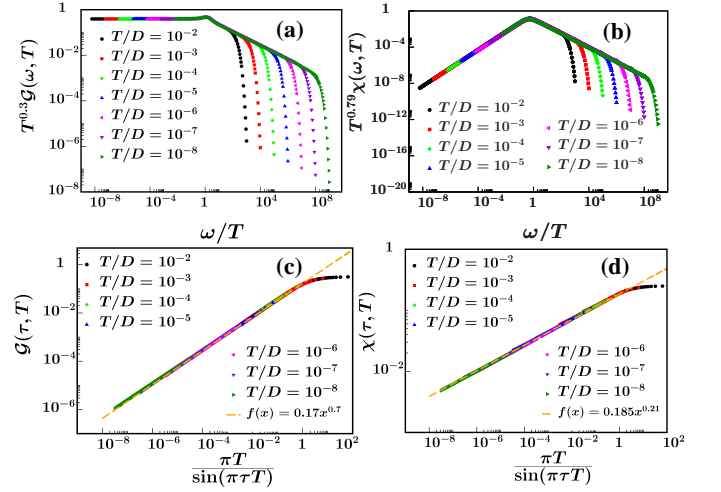}
\end{center}
\caption{Scaling functions for the imaginary part of (a) Green's function
$\mathcal{G}(\omega,T)$ and 
(b) susceptibility $\chi(\omega,T)$
 for $r=0.3$ and $\kappa=0.5$ at 
the critical $J_c\approx 1.54$ (with $T_K^0\approx 0.3D$ for $r=0$).
Both functions display $\omega/T$-scaling with scaling functions $\Phi$ obeying
$\Phi(\omega/T \longrightarrow 0) \longrightarrow c~{\rm or}~0$ 
for $\mathcal{G}$ or $\chi$,
where $c\neq 0$ is a constant.
(c),(d) the scaling functions 
%imply scaling forms
in imaginary time.
% that
%%are compatible with
% those of primary fields of 
%%a boundary conformal field theory,
%%showing power law dependence on 
%%$\pi T/\sin(\pi \tau T)$.
}
\label{fig:Figure1}
\end{figure}

By solving the large-N equations in real frequencies
for arbitrary $\omega$ and $T$~\cite{Zhu.04},
the full scaling functions in both, the quantum coherent ($\omega>T$) and 
relaxational ($T>\omega$) regimes are obtained. 
At the critical coupling $J_c(r)$, we find that all the correlators display an 
$\omega/T$-scaling.
This is demonstrated  in Fig.~\ref{fig:Figure1}(a) 
for the local singleparticle Green's function
[{\it i.e.}, the T-matrix,
${\mathcal{G}}(\omega,T)$, associated with 
${\mathcal{G}}(\tau)=G_f(\tau)G_B(\tau)$],
and in Fig.~\ref{fig:Figure1}(b) the local spin susceptibility
$\chi(\omega,T)$, which corresponds to $\chi(\tau)=-G_f(\tau)G_f(-\tau)$.
%Consequently, the QCP in the large-N limit is fully interacting.

A key insight from the large-$N$ result is that the scaling functions contain more information
beyond $\omega/T$ scaling {\it per se}. They have the particular form associated 
with a boundary conformally-invariant fixed point, depending on $\tau$ as a power 
law in  $\pi T /\sin(\pi \tau T)$~\cite{Ginsparg}.
To see this, we obtain the imaginary-time dependence
 from the real-frequency results via
\begin{equation}
\Phi(\tau)\,=\, -\eta \int_{-\infty}^{\infty}
d\omega\,
\frac{\exp(-\tau \omega)}{\exp(-\beta\omega)-\eta}
{\text{Im}} (\Phi(\omega+i0^+)) ,
\end{equation}
for $0<\tau \leq\beta$. Here, 
$\eta=\pm$ for bosonic/fermionic $\Phi$.
Figure~\ref{fig:Figure1}
 shows the (c) 
Green's function ${\mathcal{G}}(\tau,T)$ 
and (d) susceptibility $\chi(\tau,T)$ versus the combination $\pi T /\sin(\pi \tau T)$.   
Both
%, ${\mathcal{G}}(\tau,T)$ and  $\chi(\tau,T)$, 
collapse on a single scaling curve
in terms of $\pi T /(\sin(\pi \tau T))$ for 
all (low-enough) $T$.
A power-law behavior for $\tau\rightarrow 1/(2T)$ is seen 
over about 
$7$ decades,
and the exponents
are compatible with those 
for the frequency dependence.  
%%%%%%%%%%%%%%%%%%%%%%%%%%% Figure 1 %%%%%%%%%%%%%%%%%%%%%%%%%%%%%%%%%%%%%%%%%%%%
%%%%%%%%%%%%%%%%%%%%%%%%%%%%%%%%%%%%%%%%%%%%%%%%%%%%%%%%%%%%%%%%%%%%%%%%%%%%%%%

%{\it  Particle-hole symmetric pseudogap Anderson model:~}
{\it  Pseudogap Anderson model at $N=2$:~}
Guided by the large-N results, we 
%are now in a position to determine 
turn to the scaling functions for
$\mathcal{G}(\tau,T)$ and $\chi(\tau,T)$ of the
particle-hole symmetric pseudogap Anderson model at $N=2$;
the low-energy properties of this model are identical to
its pseudogap Kondo counterpart.
To this end,
we bring to bear the recently developed hybridization-expansion
Monte Carlo method~\cite{Werner.06,Rubtsov.05} on a quantum-critical model.
This 
%continuous-time quantum
%Monte Carlo approach (CT-QMC)
CT-QMC approach
%J.H. Pixley
%~\cite{Rubtsov.05} 
involves a stochastic sampling of a perturbation 
expansion in the host-impurity hybridization
or a weak coupling expansion~\cite{Prokofev.98,Rubtsov.05,Werner.06,Haule.07}. 
The results are free of any finite-size effects~\cite{Kirchner.09}.
%J. H. Pixley
%and we are able to treat the on site electron-electron interaction exactly. 
%However,
%whether this method can reach low-enough temperature with enough accuracy
%to resolve quantum critical features is an interesting question that has,
%to our knowledge, not been addressed before.

The
Anderson impurity model is defined by $\hat{H}=\hat{H}_0+\sum_{\sigma}\hat{H}_1^{(\sigma)}$
where
\bea
&&\hat{H}_0= \sum_{\kk, \sigma}
\epsilon_{\kk}\hat{n}_{\kk\sigma}
+ \sum_{\sigma}(\epsilon_d+\mbox{$\frac{1}{2}$}U\hat{n}_{d, -\sigma})\hat{n}_{d\sigma} \nonumber \\
&&\hat{H}_1^{(\sigma)}=\sum_{\kk}\left(V_{d\kk}d^{\dagger}_{\sigma}c_{\kk\sigma} + \mbox{H. c.}\right)
\label{Hamiltonian}
\eea
with $\hat{n}_{\kk\sigma}=c^{\dagger}_{\kk\sigma}c_{\kk\sigma}$, $\hat{n}_{d\sigma}=d^{\dagger}_{\sigma}d^{}_{\sigma}$,
 $\epsilon_{\kk}$ being the host dispersion, $V_{d\kk}$ the hybridization, and
$\epsilon_d$ the impurity level energy. We consider the
particle-hole symmetric case where $\epsilon_d=-\frac{1}{2}U$, with $U$ being 
the
onsite Coulomb repulsion. 
The host-impurity coupling is 
%succinctly 
specified
by the imaginary part of the hybridization function 
$\Gamma(\w)=\pi \sum_{\kk}|V_{d\kk}|^2\delta(\w-\epsilon_{\kk})$.
As in Eq.~(\ref{DOS}),
we choose
$
\Gamma(\w)=\Gamma_0\left|\frac{\w}{D}\right|^r \Theta(D-|\w|)\label{hybrid}
$.
The critical point exists only for $0< r < \frac{1}{2}$~\cite{Buxton.98}.

Central to the CT-QMC approach adopted here is 
the expansion of the partition function 
$Z=\mbox{Tr}\{\hat{T}_{\tau} e^{-\beta\hat{H}_0}\prod_{\sigma}
\mbox{exp}[-\int_0^{\beta}\mbox{d}\tau\ \hat{H}_1^{(\sigma)}(\tau)]\}$ in the 
hybridization term~\cite{Werner.06}.  
\begin{figure}[t!]
\begin{center}
\includegraphics[width=0.5\textwidth]{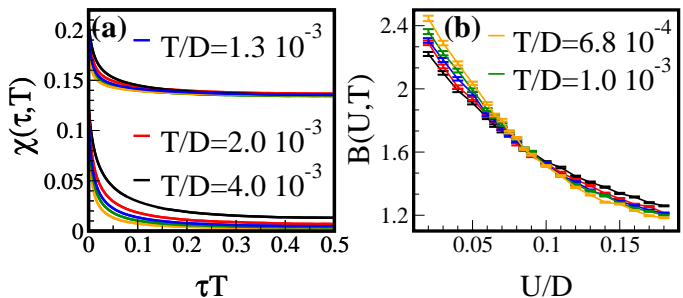}
\end{center}
\caption{Dynamical local susceptibility $\chi(\tau,T)$ 
versus $\tau T$ for $r=0.4$, and $\Gamma_0=0.1D$ with 
 $U=0$ and $U=0.3D$ (a).  Finite-temperature scaling of the Binder cumulant $B(U,T)$ as a function of $U$ at various temperatures (b), error bars are obtained from a jackknife error analysis.  From the intersection of the curves we determine the critical point to be $U_c(r=0.4)/D = 0.085 \pm 0.002$.     %For $U<U_c(r)$, 
%$\chi(\tau=1/(2T), T)$ tends to zero
%as $T$ decreases, indicating complete screening of the local moment.  
%For $U>U_c(r)$ by contrast, $\chi(\tau=1/(2T), T)$ approaches a nonzero constant
%with decreasing $T$, signaling a local moment ground state.
%At $U=0$, the exact result $\chi(\tau=0)=1/8$ is recovered.
\label{fig:Figure2}}
\end{figure}
%%%%%%%%%%%%%%%%%%%%%%%%%%%%%%%%%%%%%%%%%%%%%%%%%%%%%%%%%%%%%%%%%%%%%%%%%%%%%%%%%%%%%%%
We measure the single-particle Green's function ${\mathcal{G}}_{\sigma}(\tau)=\langle T_{\tau} d_{\sigma}(\tau)d_{\sigma}^{\dag}\rangle$, the local spin susceptibility
$\chi(\tau)=\langle T_{\tau} S_z(\tau) S_z(0)\rangle$ 
%J. H. Pixley
%with 
%
and powers of the local magnetization
$
\langle M_z^n \rangle = \langle\big( \frac{1}{\beta}\int_0^{\beta}\,d\tau S_z(\tau) \big)^n \rangle 
$
where
$S_z(\tau)=\frac{1}{2}[\hat{n}_{\uparrow}(\tau)-\hat{n}_{\downarrow}(\tau)]$.  
The static susceptibility is obtained from
$\chi(\w=0)=\int_0^{\beta}\mbox{d}\tau~\chi(\tau)$.
%In qualitative agreement with results obtained for a 
%metallic host $r=0$~\cite{Werner.07},  we find that the computation time (to achieve a fixed number of 
%measurements) for $r>0$ scales as the square of the mean perturbation order. 
%We choose a pseudogap exponent $r=0.4$ and 
%fix the strength of the hybridization $\Gamma_0=0.1$ between host and impurity.
Thermalization can be traced by $\langle n_d \rangle$ which obeys $\langle n_d \rangle=1$ in the particle-hole
symmetric model.
We also performed a binning analysis and obtained the integrated autocorrelation time which increases with 
decreasing temperature but turned out to be small (compared to the number of measurements) at all temperatures.  For the lowest temperature considered ($\beta D =9,000$) we performed $800,000$ Monte Carlo steps  for thermalization, $1,500$ Monte Carlo steps between each measurement and $18,750$ measurements.  A Monte Carlo step consists of an attempt to remove, insert and shift a segment as described in Reference~\cite{Werner.06}.
%At the lowest temperatures, we swapped up and down spin segments with equal probability, to increase the 
%sampling of the whole phase space~\cite{werner_millis.06}.
%J. H. Pixley
%In order to obtain reliable data at sufficiently low temperatures we have to be %cautious of three issues.  First, we have to ensure that measurements are taken %after the CT-QMC algorithm has converged, in order for this to be the case we %require $\langle n_d \rangle = \langle n_{d,\uparrow}\rangle + \langle %n_{d,\downarrow} \rangle \approx 1$, where $\langle n_{d,\sigma} \rangle = %\frac{1}{\beta}\langle \int_0^{\beta} n_{d,\sigma}(\tau) \mbox{d}\tau \rangle$.  %Second, we must make sure that each measurement is uncorrelated, therefore we %allow the system to evolve well past the autocorrelation time between %sequential measurements.  This becomes important for low temperatures because %we notice the autocorrelation time increases as the temperature decreases.  %Lastly, we have to be confident that the system explores all of phase space and %does not get trapped in an unphysical state.  In the two lowest temperatures %considered we use swap moves, swapping up and down spin segments 
% with equal probability, to avoid getting stuck in any region of phase %space~\cite{werner_millis.06}.
%

By varying $U$ we can tune the model through a QCP.  
%For small $U$, the local susceptibility
%approaches a constant at low temperatures --- the characteristic of a Kondo
%screened impurity moment. %In the large-$U$ phase, by contrast,
%% where the
%%impurity moment essentially decouples from the conduction band, 
%When $U$ is sufficiently large, the model is in the Kondo destroyed local moment phase, 
%and the local
%susceptibility exhibits a Curie law down to the lowest temperatures.
Correspondingly, Figure~\ref{fig:Figure2}(a) shows that
 the large-$\beta$ limit of $\chi(\tau=\beta/2,\beta)$ vanishes
for small $U$ (Kondo-screened phase) and is equal to the Curie constant for large $U$ (Kondo-destroyed local-moment phase).
%The critical value, $U_c(r)$, corresponds to where the Curie constant extrapolates
%to zero as $U$ is decreased.  
To accurately determine $U_c(r)$ we apply finite-temperature scaling to the Binder cumulant, $ B(U,T)  = \frac{\langle M_z^4 \rangle}{\langle M_z^2 \rangle^2}$,
% In analogy with classical Monte Carlo simulations 
where $1/T=\beta$ plays the role of the system size.
% and $U$ the
%classical temperature, therefore we can extrapolate our results to zero temperature to determine the location of the %QCP.  
%, similar to extrapolating classical Monte Carlo results to the thermodynamic limit to obtain the critical temperature.  
We find swap moves between up and down spin segments~\cite{werner_millis.06} are necessary to accurately measure the Binder cumulant; for the results in Figure~\ref{fig:Figure2}(b) we performed a swap move every $100$ measurements.  The nature of the intersection of the data in Figure~\ref{fig:Figure2}(b) implies that the phase transition is continuous, from the location of the intersection we obtain the critical value of $U$.  
In the quantum-critical regime the static local susceptibility displays an anomalous 
$r$-dependent exponent; we find 
\begin{equation}
\chi(T,U_c,r=0.4) \sim T^{-x},
\end{equation}
with $x=0.68(3)$  in good agreement with numerical renormalization group results~\cite{Ingersent.02}.  %{\color{red}This fractional exponent is similar to the experimental results on heavy fermion metals.} 
%As $r$ tends to zero, the exponent $x$ tends to unity from below, while, as r
%approaches 1/2 from below --- whence the critical and local-moment fixed
%points merge --- $x$ tends to 1/2.

\begin{figure}[t!]
\begin{center}
\includegraphics[width=0.5\textwidth]{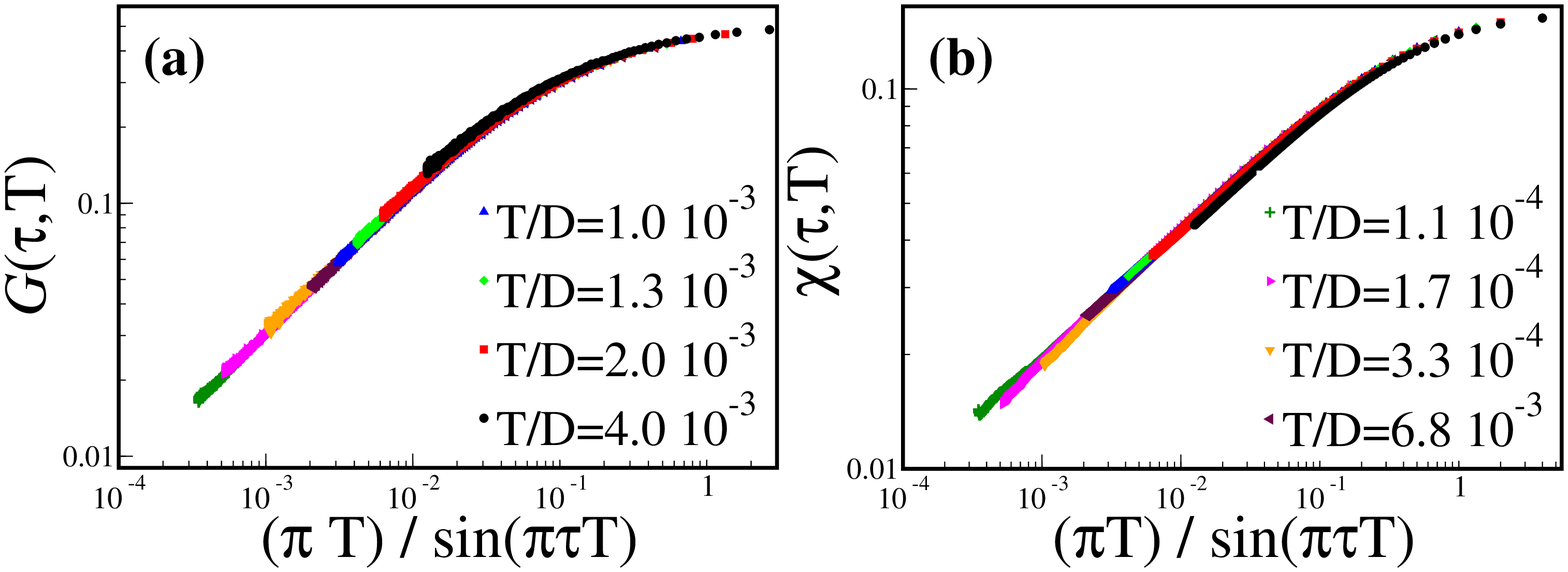}
\end{center}
\caption{Scaling of (a) Green's function
$\mathcal{G}(\tau,T)$ 
and (b) susceptibility $\chi(\tau,T)$ 
at the QCP for $r=0.4$, $\Gamma_0=0.1D$
and $U_c(r=0.4) = 0.085D$.  
The Kondo temperature in this case is $ T_K^0\approx 0.029D$ (for $r=0$).
For $T/D < 5\cdot 10^{-3}$, we observe collapse of the data  
for more than two decades of the parameter
$(\pi T)/\mbox{sin}(\pi\tau T)$, i.e., 
$\mathcal{G}_c(\tau,T)=\Psi( \pi T/\mbox{sin}(\pi\tau T))$ and
$\chi_c(\tau,T)=\Phi( \pi T/\mbox{sin}(\pi\tau T))$.  
$\Psi(y \ra 0)\propto y^\delta$ with $\delta=0.57(5)$,
and $\Phi(y \ra 0)\propto y^{1-x}$ with $x=0.68(3)$.
\label{fig:Figure3}}
\end{figure}

%We are now 
%in position to 
We now
discuss the
finite-temperature 
dynamical
scaling properties of
$\mathcal{G}(\tau,T)$ and $\chi(\tau,T)$.
Guided by the large-$N$ results, we plot 
them as functions of $(\pi T)/\mbox{sin}(\pi\tau T)$
in
Figure~\ref{fig:Figure3}.
Excellent scaling collapse is observed over
about two decades,
for all temperatures in the scaling regime.  
%The scaling exponent is compatible to the exponent $x(r)$ 
%observed in the static susceptibility.
We reach an important conclusion:
\begin{equation}
  \chi_{crit}(\tau,T)\,=\, \Phi\big(\frac{\pi\tau_0 T}
{\sin (\pi \tau
  T)}\big)\,\stackrel{T\ll T_K^{0}}{\large \sim}\,
 \big(\frac{\pi\tau_0 T}{\sin (\pi \tau
  T)}\big)^{1-x},
\label{scalingform}
\end{equation}
for $\tau^{-1}_{} \ll T_K^0$, Figure~\ref{fig:Figure3}(b).
Since $0<1-x<1$, the results for $\chi(\tau,T)$ imply that the 
order-parameter susceptibility
shows $\omega/T$-scaling.
%~\cite{note2}. 
%(%In the notation of Ref.~\cite{note1},
%here $\Delta=1-x <1$.)
A similar conclusion applies to
% the 
%local T-matrix 
%local Green's function
$\mathcal{G}(\tau,T)$,
%which,
as seen in Figure\ref{fig:Figure3}(a).  Our results yield
${\mathcal{G}}(\tau,T\rightarrow 0)\sim \tau^{-\delta}$,
with 
the exponent $\delta=1-r$, which
is believed to be exact~\cite{Fritz.06}.
%,
%also shows a
%scaling collapse in terms of  $\pi T \tau_0/\sin{(\pi \tau T)}$.
The  fact that $2 \delta \neq
1-x$ signifies the importance of vertex corrections and
in part reflects 
the interacting nature of the QCP (see below).

The boundary conformally-invariant form 
of $\chi$ and $\mathcal{G}$ immediately imply that their 
dependence on real frequency satisfies $\omega/T$ scaling
and that 
their relaxation rates,
defined in the
quantum-relaxational regime,
are linear in $T$. Expressed in terms of 
$\Gamma_M=i(\partial \ln M(\omega,T)/\partial \omega|_{\omega=0})^{-1}$
for a correlator $M$, the relaxation rates
$\Gamma_{\chi}=a
T$ and $\Gamma_{\mathcal{G}}=b T$,
where $a$ and $b$ are universal dimensionless
constants.
%follow from Eq.~(\ref{FT})  followed by analytical continuation: 
Such a linear-in-$T$ form is consistent with
what 
%happens 
has been observed 
in quantum-critical heavy
fermion compounds,
%where evidence exists for 
%the linear-in-$T$ relaxation rates
%in 
for both
the single-particle Green function~\cite{Friedemann.09} 
and order-parameter susceptibility~\cite{Schroeder.00}.
A linear-in-$T$ relaxation rate signifies that the QCP is interacting, {\it i.e.},
containing a nonzero nonlinear coupling among the critical modes. By contrast, at a Gaussian QCP (whose critical modes do not interact at the fixed point), the relaxation rate will be super-linear-in-$T$ because the nonlinear coupling itself vanishes as $T$ approaches zero~\cite{Damle.97}.

%Several remarks 
%are in order. 
%The  fact that $2 \delta \neq
%1-x$ signifies the importance of vertex corrections and
%in part reflects 
%the interacting nature of the QCP.
%This also implies that the vertex corrections preserve the scaling collapse in terms of 
%$\pi T \tau_0/\sin{(\pi \tau T)}$.

It is instructive to compare our study with previous theoretical treatments of the finite-temperature scaling behavior of the  pseudogap Anderson/Kondo model.
One study~\cite{Ingersent.02} is perturbative in %relies on perturbation in terms of 
 $r$, %the power of pseudo-gap,
 which not only becomes unreliable for finite $r$ but also does not 
allow the study of the single-particle Green's function. Another study carries out calculations in real
frequency at finite temperatures, but relies on the resummation of a perturbation series whose validity
for the quantum-critical regime is not clear~\cite{Glossop.05}. Yet another study utilizes a Callan-Symanzik approach which requires analytic continuation that is problematic as reflected in the noncommutativity
of the resummation and analytic continuation~\cite{Fritz.06}; it will be important to check whether that procedure 
yields a ${\mathcal{G}}^{'}(\omega,T)$ that is compatible in analyticity with 
${\mathcal{G}}^{''}(\omega,T)$. As a more specific illustration of our results,
we note that 
$T^r{\mathcal{G}}^{''}(\omega/T\rightarrow 0)$ is a nonzero
constant, which is contrary to both the perturbative results of Reference~\cite{Glossop.05} and 
%also in variance with 
the results
of the real-frequency Callan-Symanzik resummation
for ${\mathcal{G}}^{''}(\omega,T)$~\cite{Fritz.06}.

The scaling of the local correlators in terms of $\pi T/\sin(\pi\tau T)$
suggests that
the boundary critical state and the associated boundary operators 
may be described by their counterparts in an effective model with conformal
invariance~\cite{Kirchner.08}. 
This is so in spite of the fact that, for our problem, 
the pseudogap form of the density of states means that
the bulk fermionic component of the Hamiltonian lacks conformal 
invariance.
Hence, our results suggest an enhanced conformal symmetry that characterizes the QCP. 
%{\color{red}  Thus it appears that Kondo destroyed quantum critical points can be described by a boudnary conformal field theory.}
%%%%%%%%%%%%%%%%%%%%%%%%%%%%%%%%% FIGURE 3 %%%%%%%%%%%%%%%%%%%%%%%%%%%%%%%%%%%%%%%%%%%
%%%%%%%%%%%%%%%%%%%%%%%%%%%%%%%%%%%%%%%%%%%%%%%%%%%%%%%%%%%%%%%%%%%%%%%%%%%%%%%%%%%%%%%

\textit{Summary.}
We have obtained the full 
finite-temperature scaling functions at the local quantum-critical point of the pseudogap Anderson 
and Kondo models. 
%We have done so 
% by 
%employing 
%through 
%a  combination of techniques: perturbative RG, large-$N$ and 
%continuous-time Quantum Monte Carlo. 
Using the results directly obtained in real frequency ($\omega$) in the large-$N$ limit,
and by showing that the imaginary-time local correlators of the physical $N=2$ model
have the form of a boundary conformally invariant fixed point,
we succeeded in determining the full scaling function in 
both the quantum coherent and relaxational regimes 
%by avoiding the use of 
without using numerically ill-conditioned analytical-continuation 
schemes.
We demonstrated that the Kondo-breakdown QCP
features  a linear-in-$T$ relaxation rate for both
spin and single-electron dynamics, which is consistent with the experimental
observations in the quantum-critical heavy fermion metals.
%$\omega/T$ scaling in 
%both the 
%%local T-matrix 
%single-particle Green's function and the order parameter susceptibility.
%As a by-product, we showed that the 
%recently developed continuous-time QMC for fermions~\cite{Rubtsov.05,Werner.06} 
%can access sufficiently low temperatures even in a 
%quantum critical system, so much so that scaling functions 
%for dynamical quantities can be reliably determined.\\

We thank L.~Fritz, K.~Ingersent, M.~Vojta and P.~Werner
for useful discussions.
This work has been supported in part by
NSF (Grant No. DMR-1006985),
the Robert A. Welch Foundation (Grant No. C-1411),
the W. M. Keck Foundation,
and the Rice Computational Research Cluster
funded by NSF.
%%%%%%%%%%%%%%%%%%%%%%%%%%%%%%%%%%%%%%%%%%%%%%%%%%%%%%%%%%%%%%%%%%%%%%%%%%%%%%%%%%%%%%

\end{document}